\documentclass[journal]{IEEEtran}

\ifCLASSINFOpdf
\else
 \usepackage[dvips]{graphicx}
\fi

\usepackage[cmex10]{amsmath}
\usepackage{multirow}   
\usepackage{subfigure}
\usepackage[justification=centering]{caption}
\usepackage{multicol}

\begin{document}

\title{
Fractional Order Modeling of Human Operator Behavior with Second Order Controlled Plant and Experiment Research }

\author{Jiacai Huang,
        ~Yangquan Chen$^{\ast}$\thanks{$^\ast$Corresponding author. Email: ychen53@ucmerced.edu},~\IEEEmembership{Senior member,~IEEE,}
        ~Haibin Li,
        ~Xinxin Shi,

\thanks{Jiacai Huang, Haibin Li and Xinxin Shi are with School of Automation, Nanjing Institute of Technology, Nanjing 211167, China. e-mail: huangjiacai@126.com}%

\thanks{Yangquan Chen is with School of Engineering, University of California, CA 95343, USA. e-mail: ychen53@ucmerced.edu}%

\thanks{*This work was supported by National Natural Science Foundation (NNSF) of China under Grant(61104085), Natural Science Foundation of Jiangsu Province (BK20151463, BK20130744), Innovation Foundation of NJIT(CKJA201409,CKJB201209), sponsored by Jiangsu Qing Lan Project, and the Jiangsu Government Scholarship for Overseas Studies(JS-2012-051)}

\thanks{Manuscript received August 31, 2015; revised ***, ***.}}

\markboth{IEEE/CAA JOURNAL OF AUTOMATICA SINICA,~Vol.~X, No.~X, X~X}%
{Shell \MakeLowercase{\textit{et al.}}: Bare Demo of IEEEtran.cls
for Journals}

\maketitle

\begin{abstract}
Modeling human operator's dynamic plays a very important role in the manual closed-loop control system, and it is an active research area for several decades. Based on the characteristics of human brain and behaviour, a new kind of fractional order mathematical model for human operator in SISO systems is proposed. Compared with the traditional models based on the commonly used quasi-linear transfer function method or the optimal control theory method, the proposed fractional order model has simple structure with only few parameters, and each parameter has explicit physical meanings. The actual data and experiment results with the second-order controlled element illustrate the effectiveness of the proposed method.

\end{abstract}


\IEEEpeerreviewmaketitle

\section{Introduction}

\IEEEPARstart{T}{he} modeling of human operator is still an open problem. In manual closed-loop control system, the accurate mathematical model of human operator is very important and provides criteria to the controller design of the manual control system. The human operator is a very complex system whose behaviour range includes not only skilled control tasks, but also instinctive and emotional reactions, such as those resulting from pain or fear.

For decades, modeling human operator's dynamic has been an active research area.The earliest study that considered the human operator as a linear servomechanism is Tustin in 1947\cite{c1}, who proposed that the main part of the operator's behaviour might be described by an 'appropriate linear law', despite the amplitude nonlinear variations and haphazard fluctuations. In 1948, reference \cite{c2} studied the human operator as an engineering system, and proposed the following theory of the human operator in control system: the human operator behaves basically as an intermittent correction servo which consists of ballistic movement, moreover there are some counteracting processing tending to make controls seem continuous. In 1959, McRUER considered the role of human elements in certain closed loop control systems and proposed a quasi-linear mathematical model for the human operator, and the proposed model is composed of two components-a describing function and remnant \cite{c3}. In \cite{c4}, the rms-error performance of a human operator in a simple closed-loop control system was measured and compared with the performance of an `optimum' linear controller, the comparison results showed that the human operator perform about as well as a highly constrained optimum linear controller. In \cite{c5} the human operators were considered as a monitor and controller of multidegree of freedom system, and the experiment results showed that the human operators are in fact random sampling device and nearly ideal observers, meanwhile individual operator may have fixed patterns of scanning for a short periods and change the patterns from time to time, and different human operators have different patterns.

In 1965, McRUER\cite{c6} studied the human pilot dynamics in compensatory system and proposed human pilot models with different controlled element, and the experiments results validated the proposed models. In 1967, McRUER summarized the current state of the quasi-linear pilot models, including experimental data and equations of describing function models for compensatory, pursuit, periodic, and multiloop control situations \cite{c7}. In \cite{c8}, the deficiencies of the existing quasi-linear pilot models have been analyzed and then some new analytical approaches from automatic control theory have been proposed to estimate pilot response characteristics for novel situations.

In \cite{c12}, based on the assumption that the operator behaves as an optimal controller and information processor subject to the operator¡¯s inherent physical limitations, a mathematical model of the instrument-monitoring behavior of the human operator was developed. In \cite{c14}, an adaptive model with variable structure was presented to describe the behavior of the human operator in response to sudden changes in plant dynamics and transient disturbances. In \cite{c15}, a pilot model based on Kalman filtering and optimal control was given which provides for estimation of the plant state variables, the forcing functions, the time delay, and the neuromuscular lag. The remnant which is an important component of the quasi-linear model for the human operator was discussed in in \cite{c16}, and a model for remnant was postulated in which remnant is considered to arise from an equivalent observation noise vector whose components are linearly independent white noise processes. In \cite{c17} and \cite{c18}, a mathematical model of the human as a feedback controller was developed using optimal control and estimation theory.

From 70s to the early 21st century, the problem of human operator modeling has been widely studied and a lot of new achievements emerged [15-28].

In recent years, with the new situation and different application, the modeling human operator's dynamic is still an active research area. In \cite{c71}, a two-step method using wavelets and a windowed maximum likelihood estimation method was proposed for the estimation of a time-varying pilot model parameters. In \cite{c88} the human control model in teleoperation rendezvous on the basis of human information processing was studied, and the longitudinal and lateral control models for the human operator were presented based on phase plane control method and fuzzy control method. In \cite{c89}, a review of pilot model used for flight control system design that focuses specifically on physiological and manual control aspects was presented.

For a human-in-the-loop system in safety-critical application, the correctness of such systems depends not only on the autonomous controller, but also on the actions of the human controller. In \cite{c91} a formalism for human-in-the-loop control systems was presented which focuses on the problem of synthesizing a semi-autonomous controller from high-level temporal specification that expect occasional human intervention for correct operation. In \cite{c95} the three different approaches (Engineering, Physiology and Applied Experimental Psychology) to the study of human operator have been discussed, and the importance of the studying the human operator has been pointed out. In \cite{c96} the accurate control of human arm movement in machine-human cooperative control of GTAW process was studied and an adaptive ANFIS model was proposed to model the intrinsic nonlinear and time-varying characteristic of the human welder response, at last the human control experimental results verified that the proposed controller was able to track varying set-points and is robust under measurement and input disturbances.

The existing models for human operator are complicated and established by integer order calculus. In this paper, based on the characteristics of human brain and behaviour, the fractional order human operator model is proposed and validated by the actual data.



\section{Fractional Order Calculus}
Fractional calculus has been known since the development of the integer order calculus, but for a long time it has been considered as a sole mathematical problem. In recent decades, fractional calculus has become an interesting topic among system analysis and control fields due to its long memory characteristic \cite{Chen2009}, \cite{Chen2010}, \cite{Oldham1974}, \cite{Podlubny1999}, \cite{Li2010}, \cite{Podlubny1999-2}.

Fractional calculus is a generalization of integer order integration and differentiation to non-integer order ones.Let symbol $_aD_t^\lambda$ denotes the fraction order fundamental operator, defined as follows \cite{Chen2009}:
\begin{equation}
\label{eq1}
{}_aD_t^\lambda  \buildrel \Delta \over = {D^\lambda } = \left\{ \begin{array}{ll}
\frac{d^\lambda}{d{t^\lambda }},~~~~& R(\lambda ) > 0;\\\\
1,~~~& R(\lambda ) = 0;\\\\
\int_a^t {(d\tau )}^{ - \lambda },~~~& R(\lambda ) < 0.
\end{array} \right.
\end{equation}
Where $a$  and $t$  are the limits of the operation, $\lambda$  is the order of the operation, and generally $\lambda\in R$ and $\lambda$ can be a complex number.

The three most used definitions for the general fractional differentiation and integration are the Grunwald-Letnikov (GL) definition \cite{Chen2010}, the Riemann-Liouville (RL) definition and the Caputo definition \cite{Oldham1974}.

The GL definition is given as \begin{equation}
\label{eq2}
{}_aD_t^\lambda f(t) = \mathop {\lim }\limits_{h \to 0} {h^{ - \lambda }}\sum\limits_{j = 0}^{\left[ {\frac{{t - a}}{h}} \right]} {{{( - 1)}^j}} \left( \begin{array}{l}
\lambda \\
j
\end{array} \right)f(t - jh)
\end{equation}
where $[\cdot]$ means the integer part, $h$ is the calculus step, and $\left( \begin{array}{l}
\lambda \\
j
\end{array} \right) = \frac{{\lambda !}}{{j!\left( {\lambda  - j} \right)!}}$
 is the binomial coefficient.

The RL definition is given as
\begin{equation}
\label{eq3}
{}_aD_t^\lambda f(t) = \frac{1}{{\Gamma (n - \lambda )}}\frac{{{d^n}}}{{d{t^n}}}\int_a^t {\frac{{f(\tau )}}{{{{(t - \tau )}^{\lambda  - n + 1}}}}d\tau }
\end{equation}
where $n-1<\lambda< n$ and $\Gamma(\cdot)$ is the Gamma function.

The Caputo definition is given as
\begin{equation}
\label{eq4}
{}_aD_t^\lambda f(t) = \frac{1}{{\Gamma (n - \lambda )}}\int_a^t {\frac{{{f^n}(\tau )}}{{{{(t - \tau )}^{\lambda  - n + 1}}}}d\tau }
\end{equation}
where $n-1<\lambda<n$.

Having zero initial conditions, the Laplace transformation of the RL definition for a fractional order $\lambda$ is given by
$L\left\{ {{}_aD_t^\lambda f(t)} \right\} = {s^\lambda }F(s)$,
where $F(s)$ is Laplace transformation of $f(t)$.

\section{Review of the Quasi-linear Models for Human Operator}
The quasi-linear transfer function is an effective method for the modeling of human operator, and the quasi-linear models have been found to be useful for the analysis of closed loop compensatory behaviour in the manual control system. For a simple compensatory manual control system, the functional block  diagram is shown as Fig.1, where $i(t)$ is the system input, $e(t)$ is the system error, $c(t)$ is the human operator output, $m(t)$ is the system output.
\begin{figure}[!htb]
  \centering
  \includegraphics[width=0.9\hsize]{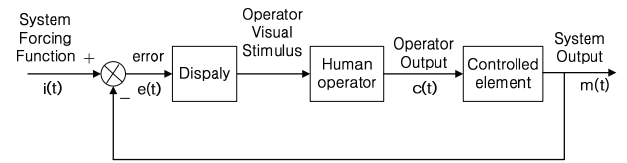}
  \caption{Functional block diagram of the manual control system}
  \label{fig1}
\end{figure}

For the above compensatory manual control system, the generalized form of the quasi-linear model for human operator was proposed as follow \cite{c3,c6,c7,c8}:
\begin{equation}
\label{eq5}
{Y_{P1}}(s) =\frac{C(s)}{E(s)} =K_p{\frac{{\tau_L}s+1}{{\tau_I}s+1}}{\frac{e^{{-L}s}}{{\tau_N}s+1}}
\end{equation}
Where $C(s)$ and $E(s)$ are the Laplace transform of $c(t)$ and $e(t)$ respectively, ${\tau_L}$ and ${\tau_I}$  represent the equalisation characteristics of human operator, $L$ and ${\tau_N}$ represent the reaction time and neuromuscular delay of human operator respectively, ${K_p}$ represents the human operator¡¯s gain which is dependant on the task and the operator¡¯s adaptive ability. The parameters in the above transfer function are adjustable as needed to make the system output follow the forcing function, i.e., the parameters, as adjusted, reflect the operator¡¯s efforts to make the overall system (including himself) stable and the error small. The quasi-linear model of Eq.(\ref{eq5}) has been widely quoted by further research.

\begin{figure}[!htb]
  \centering
  \includegraphics[width=0.9\hsize]{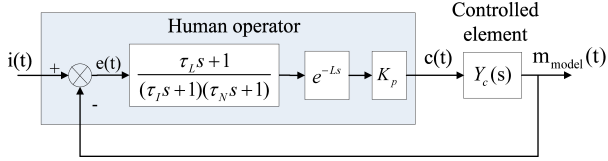}
  \caption{The mathematical model of the manual control system}
  \label{fig2}
\end{figure}

Based on the human operator model described by Eq.(\ref{eq5}), the mathematical model of the manual control system is shown in Fig.\ref{fig2}.

In reference \cite{c22}, a detail research was made to the compensatory manual control system which as shown in Fig.1, in which the forcing function (i.e. the system input) $i(t)$ is a random appearing signal, and in the human operating process, the error $e(t)$ and human output $c(t)$ can be obtained. By studying the relationship between the error $e(t)$ and human output $c(t)$, the mathematical models for human operator with respect to controlled elements was proposed \cite{c22}.

\emph{\textbf{Because the second order controlled element includes not only those which reflect the particular nature, but also represent the classic and representative about model, in this paper we take second order controlled element as as example, which is described as follow:}}
\begin{equation}
\label{eq6}
{Y_c}(s) ={\frac{K_c}{s(Ts+1)}}, ~~~~~T=\frac{1}{3}, K_c=1,
\end{equation}
then the system input $i(t)$, the system output $m(t)$, the system error $e(t)$, the human operator output $c(t)$ and the lag of the operator output $\frac{C(s)}{s+3}$ were recorded as Fig.3(a)-Fig.3(e).
\begin{figure}[!htbp]
  \centering
  \subfigure[System input $i(t)$]{
  \label{fig:subfig:a} 
  \includegraphics[width=0.9\hsize,height=0.6\hsize]{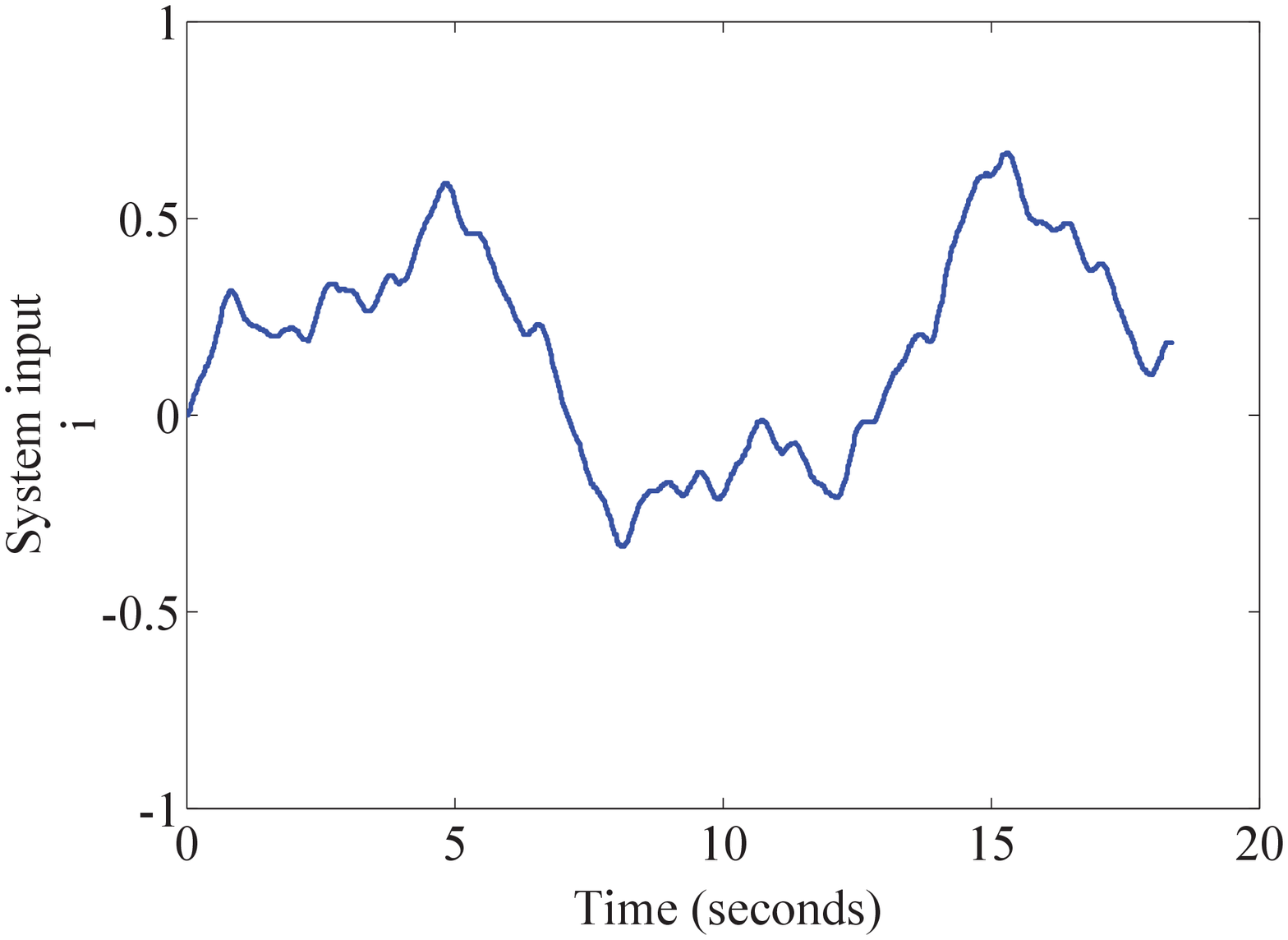}}
\end{figure}
\begin{figure}[!htbp]
  \centering
    \subfigure[System output $m(t)$]{
    \label{fig:subfig:b} 
    \includegraphics[width=0.9\hsize,height=0.6\hsize]{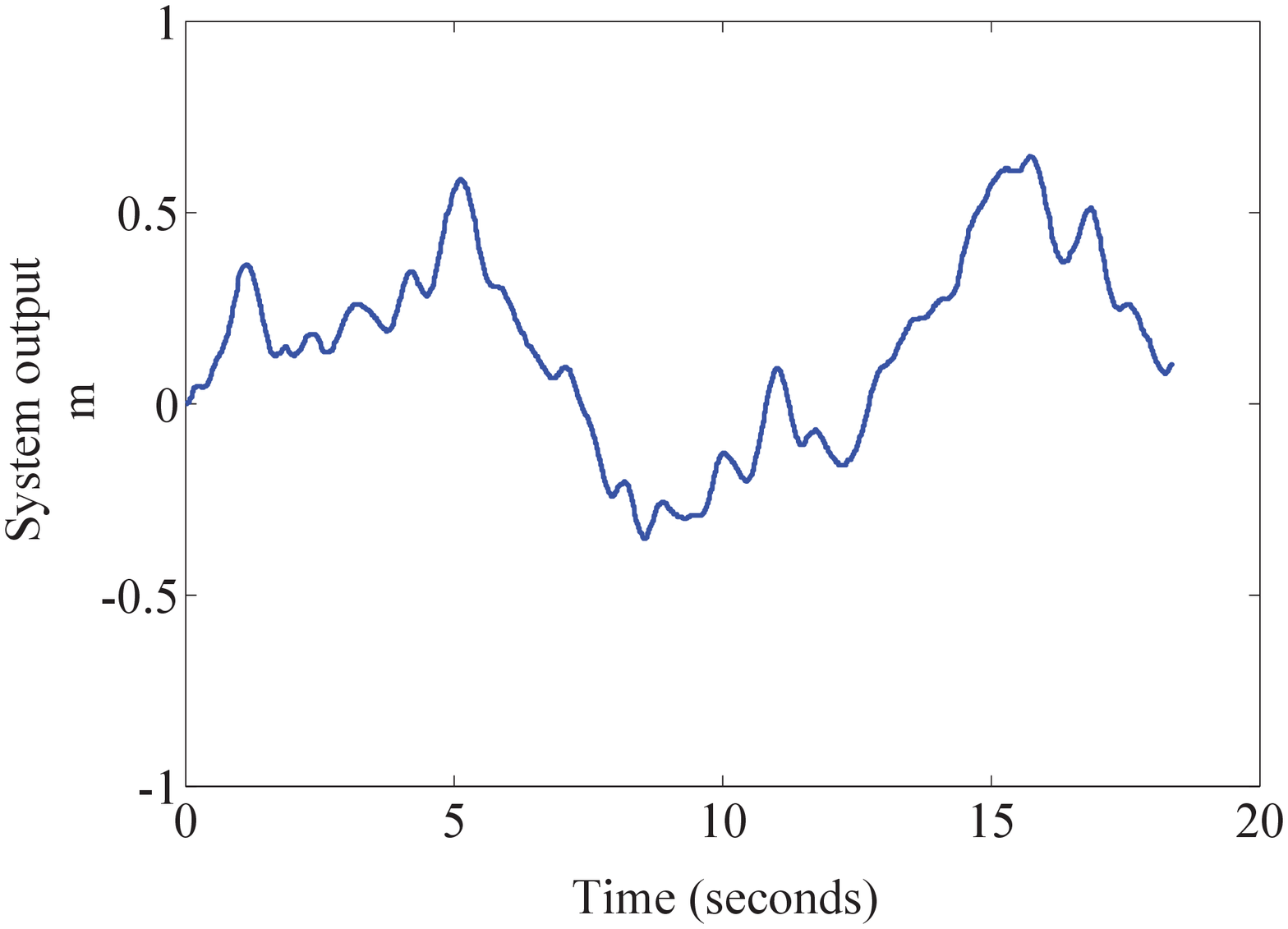}}
\end{figure}
\begin{figure}[!htbp]
  \centering
  \subfigure[System error $e(t)$]{
  \label{fig:subfig:c} 
  \includegraphics[width=0.9\hsize,height=0.6\hsize]{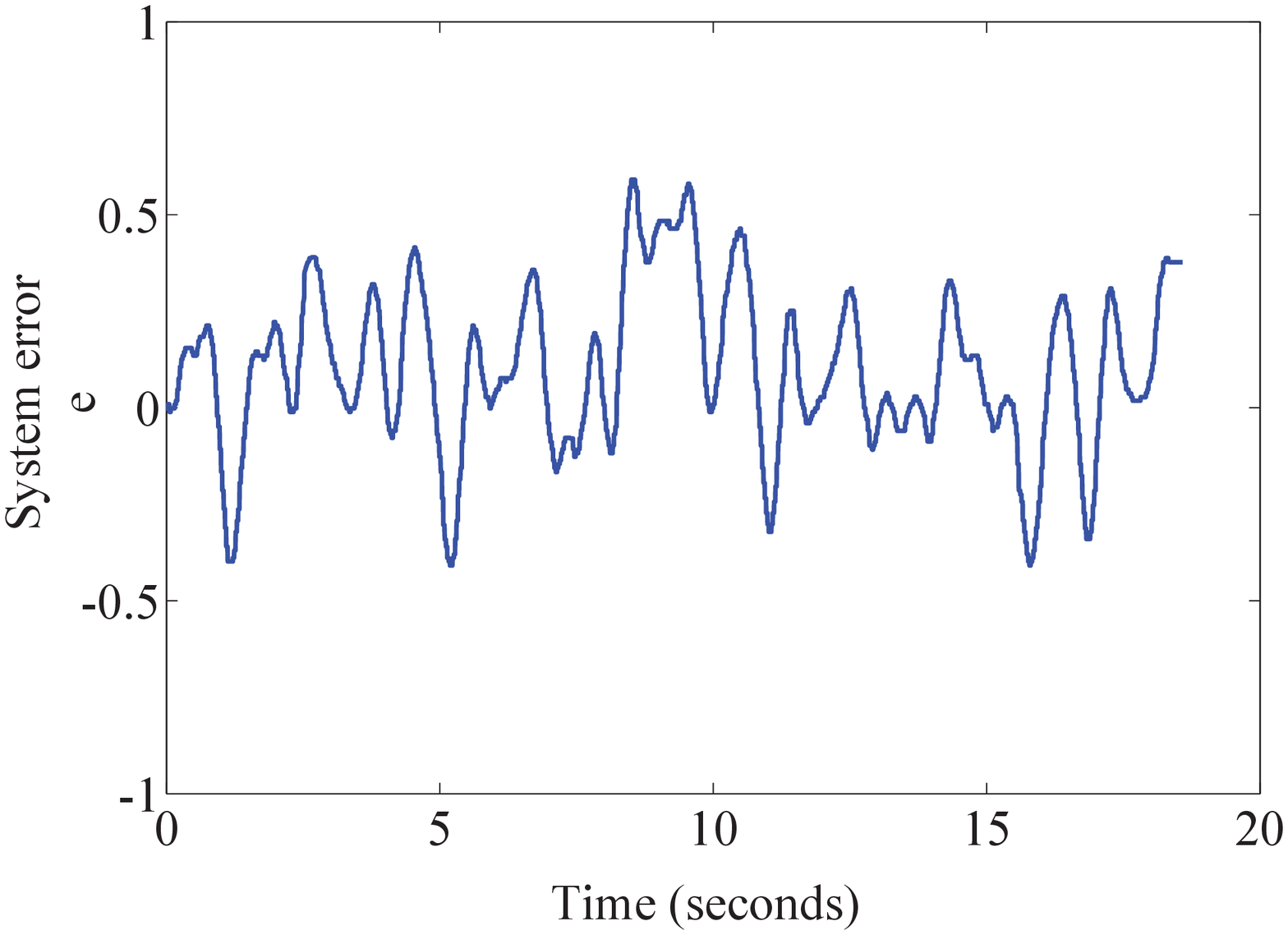}}
\end{figure}
\begin{figure}[!htbp]
  \centering
   \subfigure[Human operator output $c(t)$]{
    \label{fig:subfig:d} 
    \includegraphics[width=0.9\hsize,height=0.6\hsize]{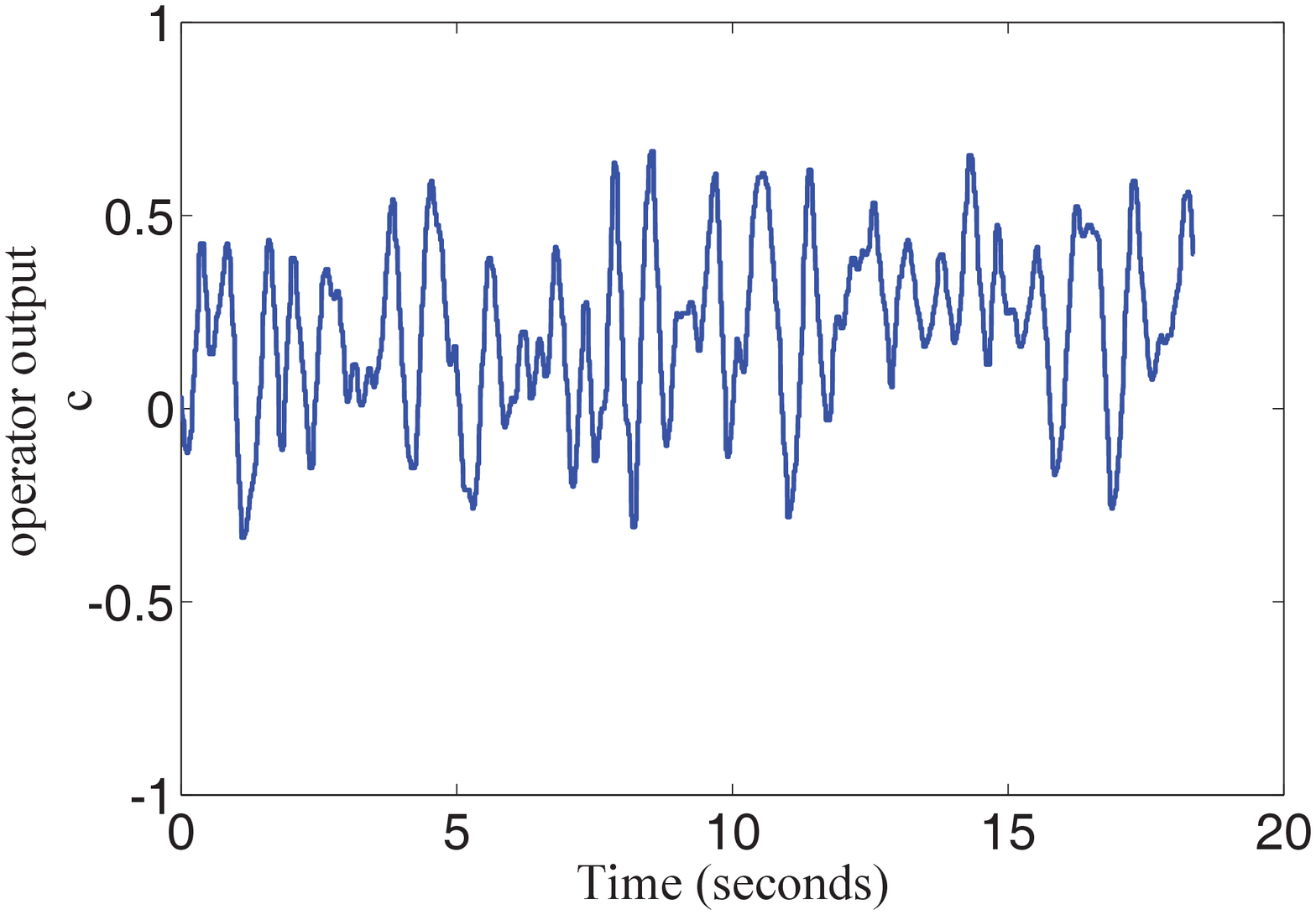}}
  \end{figure}
\begin{figure}[!htbp]
  \centering
     \subfigure[ The lag of operator output, i.e. $\frac{c}{s+3}$]{
    \label{fig:subfig:e} 
    \includegraphics[width=0.9\hsize,height=0.6\hsize]{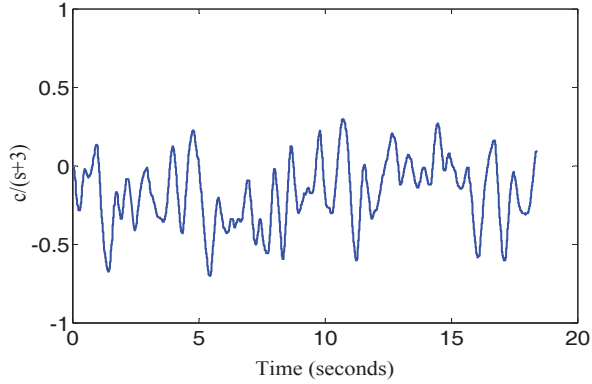}}
    \caption{ Manual control system response, ${Y_c}(s) =\frac{K_c}{s(Ts+1)}$,with $T=\frac{1}{3}$,$K_c=1$}
  \label{fig3} 
\end{figure}
From the above experiment result, when the lag of the operator output c(t) (i.e. fig.3(d))is compared with the system error e(t), a great similarity can be seen, so the following transfer function between c(t) and e(t) was proposed in \cite{c22} :

\begin{equation}
\label{eq7}
{Y_{P2}}(s) = \frac{{C(s)}}{{E(s)}} = K_p(s+\frac{1}{T})e^{-Ls}=K_p(s+3)e^{-Ls},
\end{equation}
where $K_p$ is the human operator's gain; $L$ is the time delay of human operator, which is about L{\rm{=}}0.16s.

\begin{figure}[!htb]
  \centering
  \includegraphics[width=0.9\hsize]{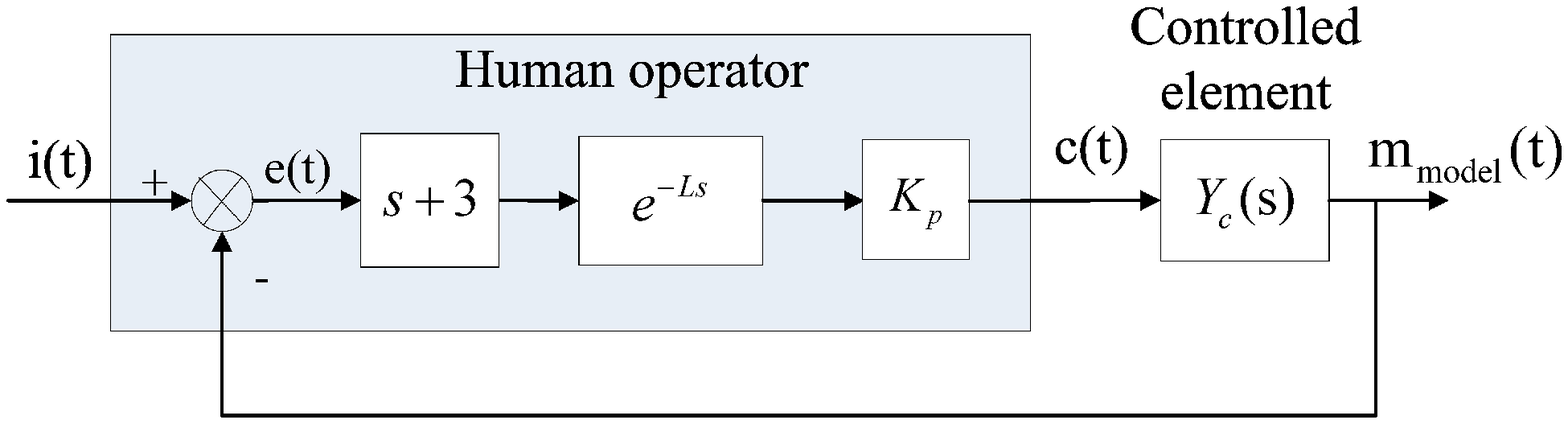}
  \caption{The mathematical model of the manual control system}
  \label{fig4}
\end{figure}

Based on the human operator model described by Eq.(\ref{eq7}), the mathematical model of the manual control system is shown in Fig.\ref{fig4}.

\section{Fractional Order Mathematical Model For Human Operator Behavior}
In the existing research, the human operator models are established based on the integer order calculus. In fact, the human body is a high nonlinear servomechanism, the control task is completed through the cooperation of the eyes, the brain/nervous system, the muscle and the hands, as shown in Fig.5.
\begin{figure}[!htb]
  \centering
  \includegraphics[width=0.6\hsize]{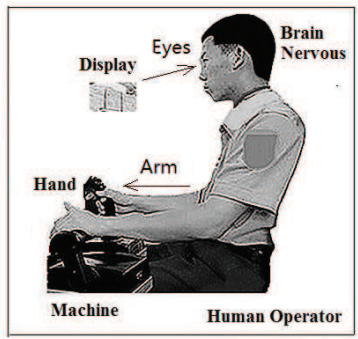}
  \caption{The control structure of a human operator}
  \label{fig5}
\end{figure}

Let us consider the manual control system shown in Fig.1, in which the human operator is shown in fig.3. In this system, the human operator controls the machine by hands to follow the target. The eyes act as a sensor, the brain acts as controller and sends the nervous system signal to the arm and hand to follow the target. The muscles of the arm and hand are employed as power actuators. Meanwhile the human has the following characteristic\cite{c1,c91}:

~(1) \emph{\textbf{For human brain, the later the thing happens, the clearer the memory is. On the contrary, the earlier, the poorer. In other words, the human brain has higher memory level for the newer things, and lower memory level for the older things.}}

~(2) During the human action, there exist dead-time in the nervous system, including the dead-time from the retina to the brain, and the dead time from the brain to the muscle.

~(3) The human muscle has the viscoelastic property.

From the above facts, it can be concluded that the dynamics of the human operator¡¯s brain is most like a kind of fractional order integral or derivative which exhibits a long memory characteristics, and so the human operator can be seen as a fractional order controller with time delay, \emph{\textbf{then in this paper the fractional order model for human operator in SISO systems is proposed as follow:}}

\begin{equation}
\label{eq8}
Y_{P3}(s) = \frac{{C(s)}}{{E(s)}} = \frac{{K_p e^{- Ls}}}{{s^\alpha }},\alpha \in R.
\end{equation}

Where $\alpha$ is the fractional order which describes the dynamics of the human operator, and $\alpha$ can be positive or negative; $K_p$ is the human operator's gain; $L$ is the total time delay of human operator, including the dead-time in the nervous system from the retina to the brain, and the dead time in the nervous system from the brain to the muscle. \emph{\textbf{In real system, the $\alpha$ and other parameters can be obtained by online or off-line identification.}}

Based on fractional order model of the human operator described by Eq.(\ref{eq8}), the mathematical model of the manual control system is shown in Fig.6.
\begin{figure}[!htb]
  \centering
  \includegraphics[width=0.9\hsize]{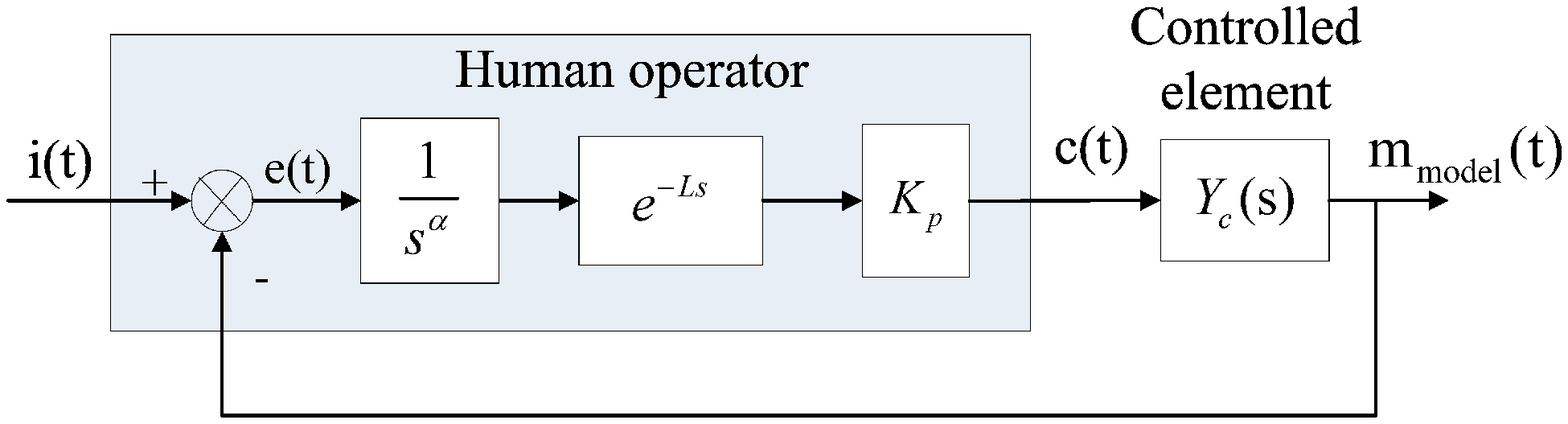}
  \caption{The mathematical model of the manual control system}
  \label{fig6}
\end{figure}

In the following section, the effectiveness of the proposed fractional order model for human operator will be validated.

\section{Model Validation with Actual Data }

In this section, the off-line verification and comparison will be done to the traditional mathematical models described by Eq.(\ref{eq5}) and Eq.(\ref{eq7}), and the new proposed fractional order model described by Eq.(\ref{eq8}). In the model verification process, the $best\_fit$ parameters for the above three models have been obtained by the $fminsearch$ function with actual data taken from reference \cite{c22}, and the following cost function, i.e. the root mean square error(RMSE) is used,
\begin{equation}
\label{eq9}
J = \sqrt {\frac{{\int_0^T {\left( {m_{{\rm{model}}} (t) - m(t)} \right)^2 } }}{T}},
\end{equation}
where $m(t)$ is the actual output of the manual control system, $m_{model}(t)$ is the model output of the manual control system by using the human operator model and the actual input (i.e. as shown in Fig.(\ref{fig2}),Fig.(\ref{fig4}) and Fig.(\ref{fig6})), $T$ is the operating time period of human operator.

In order to get the $best\_fit$ parameters of each model, the following searching criteria is adopted.

Case 1: When the human operator model is described by Eq.(\ref{eq8}), i.e. the proposed fractional order model, the searching criteria is
\begin{equation}
\label{eq10}
\left\{ {\alpha^* ,K_p^* ,L^*} \right\}_{best\_fit}  = \mathop {{\mathop{\rm mi}\nolimits} {\rm{n}}}\limits_{\alpha  \in R;K_p ,L \in R^ + } (J).
\end{equation}

In this case, the fractional order differentiation/integration symbol $\frac{1}{s^\alpha}$ is implemented by the Grunwald-Letnikov(GL) definition described as Eq.(\ref{eq2}).

Case 2: When the human operator model is described as Eq.(\ref{eq7}), i.e. the traditional model, the searching criteria is
\begin{equation}
\label{eq11}
\left\{ {K_p^* ,L^*} \right\}_{best\_fit}  = \mathop {{\mathop{\rm mi}\nolimits} {\rm{n}}}\limits_{K_p, L \in R^+} (J).
\end{equation}

Case 3: When the human operator model is described as Eq.(\ref{eq5}), i.e. the traditional model, the searching criteria is
\begin{equation}
\label{eq12}
\left\{ {T_L^* ,T_I^* ,T_N^* ,K_p^* ,L^* } \right\}_{best\_fit}  = \mathop {{\mathop{\rm mi}\nolimits} {\rm{n}}}\limits_{T_L ,T_I ,T_N ,K_p ,L \in R^ +  } (J).
\end{equation}

\subsection{ The minimum RMSE and $best\_fit$ parameters for each models}

Using the above searching criteria Eq.(\ref{eq10})$-$  Eq.(\ref{eq12}), the minimum RMSE and the corresponding $best\_fit$ parameters value for each model are obtained as shown in Table 1. From Table 1, it is obvious that the proposed fractional order model described by Eq.(\ref{eq8}) has the smallest RMSE, and the corresponding order of the model is $\alpha=-0.4101$. This means that compared with the traditional model, the proposed fractional order model described by Eq.(\ref{eq8}) is the $best\_fit$ model for describing the human operator behavior, in other word, the human operator is a fractional order system.
\begin{table}[h]
\caption{$best\_fit$ parameters value and RMSE for each model}
\label{table_1}
\begin{center}
\begin{tabular}{|c|l|c|}

\hline
Model                                                                                                 & Parameters   & Values  \\ \cline{2-3}

\hline
\multirow{4}*{$Y_{P3}(s){\rm{ = }}\frac{{K_p e^{ - Ls} }}{{s^\alpha  }}$}                           & RMSE         & 0.0012 \\ \cline{2-3}
                                                                                                     & $\alpha^*$     &-0.4101 \\ \cline{2-3}
                                                                                                     & $K_p^*$        & 4.403  \\ \cline{2-3}
                                                                                                     & $L^* (sec)$    & 0.117  \\ \cline{2-3}
\hline
\multirow{3}*{$Y_{P2}(s){\rm{ = }}K_p(s+3)e^{-Ls}$}                                                     & RMSE         & 0.0018  \\ \cline{2-3}
                                                                                                     & $K_p^*$        & 7.994   \\ \cline{2-3}
                                                                                                     & $L^* (sec)$    & 0.014   \\ \cline{2-3}
\hline
\multirow{6}*{$Y_{P1} {\rm{(}}s{\rm{)=}}\frac{{K_p (T_L s + 1)e^{ - Ls} }}{{(T_I s + 1)(T_N s + 1)}}$}  & RMSE          & 0.0024  \\ \cline{2-3}
                                                                                                     & $K_p^*$         & 1.7298    \\ \cline{2-3}
                                                                                                     & $T_L^*$         & 1.8146    \\ \cline{2-3}
                                                                                                     & $T_I^*$         & 0.162     \\ \cline{2-3}
                                                                                                     & $T_N^*$         & 0.162     \\ \cline{2-3}
                                                                                                     & $L^* (sec)$     & 0.006    \\ \cline{2-3}
\hline
\end{tabular}
\end{center}
\end{table}

\subsection{The RMSE of the proposed fractional order model to different $\alpha$ and $L$ }
\begin{figure}[!htb]
  \centering
  \includegraphics[width=0.9\hsize]{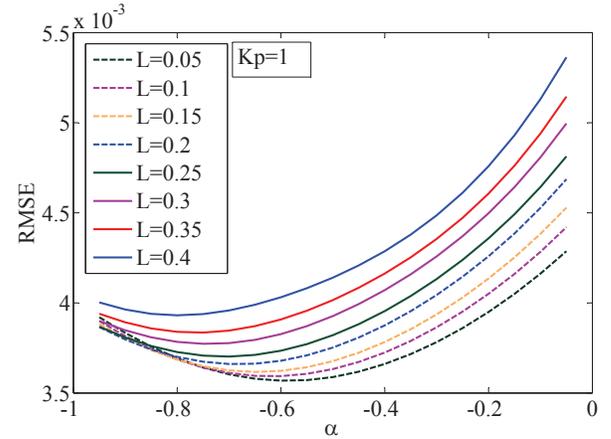}
  \caption{The RMSE scan result to different $\alpha$ with fixed $L$, and $K_p=1$}
  \label{fig7}
\end{figure}

In this section, the RMSE of the proposed model described by Eq.(\ref{eq8}) to different $\alpha$, $L$ and $K_p$ will be scanned. Because the time delay and gain of human operator have finite range, so in this scanning process, the time delay $L$ gets some fixed value between $0$ to $0.4$, and the gain $K_p$ gets the fixed value of $1$, $3$ and $5$. For each $K_p$ and $L$, the $\alpha$ is varied from $-0.95$ to $-0.05$ with $0.05$ step length. The scan results are shown in Fig.(\ref{fig7}) to Fig.(\ref{fig11}).

\begin{figure}[!htb]
  \centering
  \includegraphics[width=0.9\hsize]{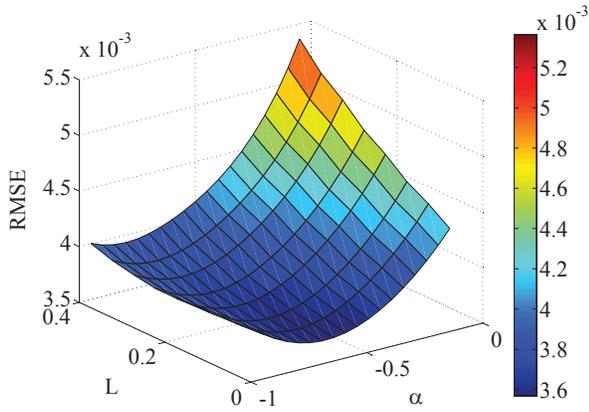}
  \caption{The 3-D RMSE scan result to different $\alpha$ and $L$, and $K_p=1$}
  \label{fig8}
\end{figure}

~~~(1)~ When the gain of the human operator is $K_p=1$, the RMSE scan result for each $L$ is shown in Fig.(\ref{fig7}), and the 3-D RMSE scan result to different $\alpha$ and $L$ is shown in Fig.(\ref{fig8}). From Fig.(\ref{fig7}) and Fig.(\ref{fig8}) it is clear that:(a) the corresponding $\alpha$ to the minimum RMSE is fractional; (b)when the time delay $L$ gets bigger valve, the corresponding minimum RMSE is also bigger.

\begin{figure}[!htb]
  \centering
  \includegraphics[width=0.9\hsize]{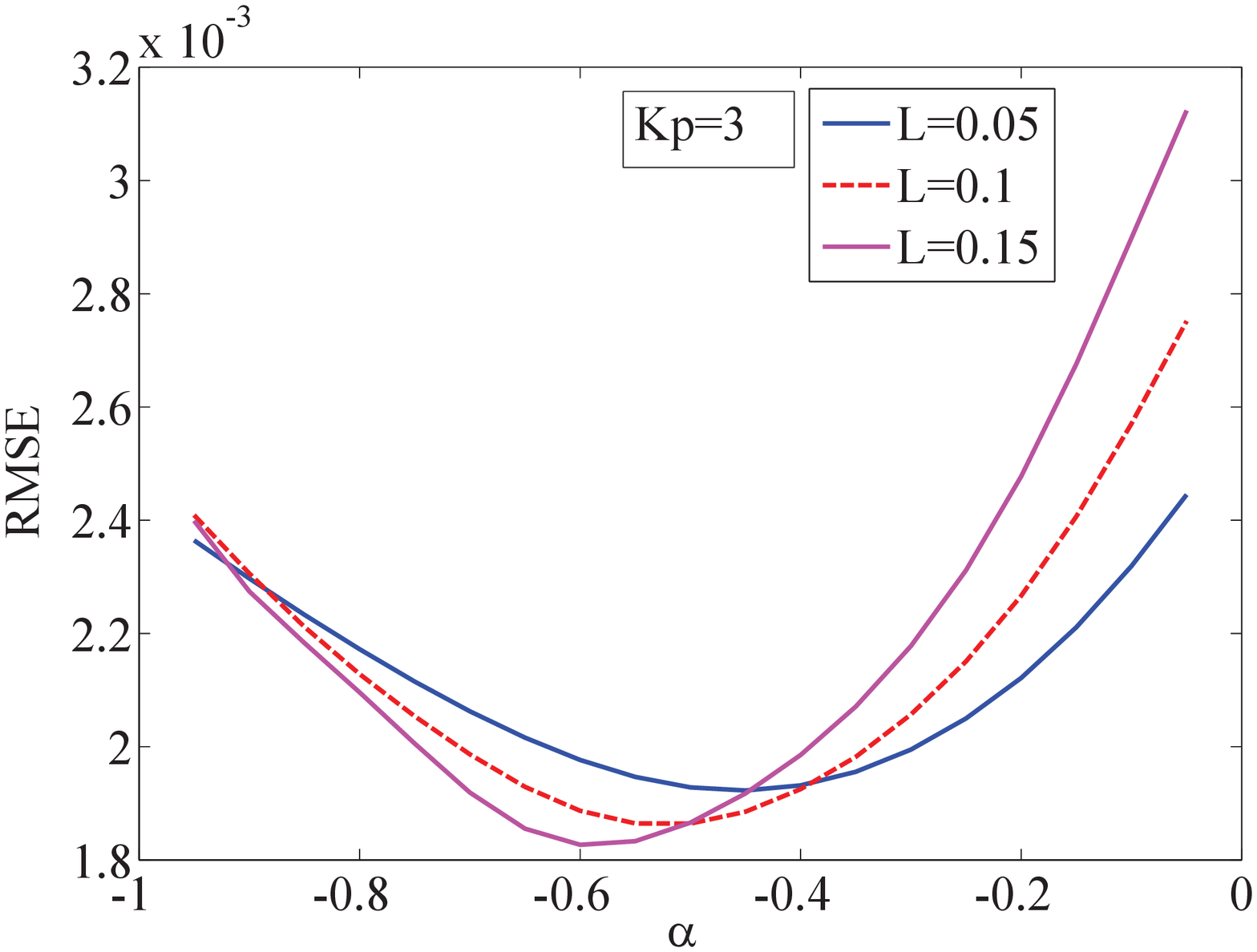}
  \caption{The RMSE scan result to different $\alpha$ with fixed $L$, and $K_p=3$}
  \label{fig9}
\end{figure}

~~~(2)~ When the gain of the human operator is $K_p=3$, the RMSE scan result for each $L$ is shown in Fig.(\ref{fig9}),from which it can be seen that: (a)the corresponding $\alpha$ to the minimum RMSE is fractional; (b)when the time delay $L$ gets smaller valve, the corresponding minimum RMSE is bigger, this is because the human gain gets the bigger value in this case.

\begin{figure}[!htb]
  \centering
  \includegraphics[width=0.9\hsize]{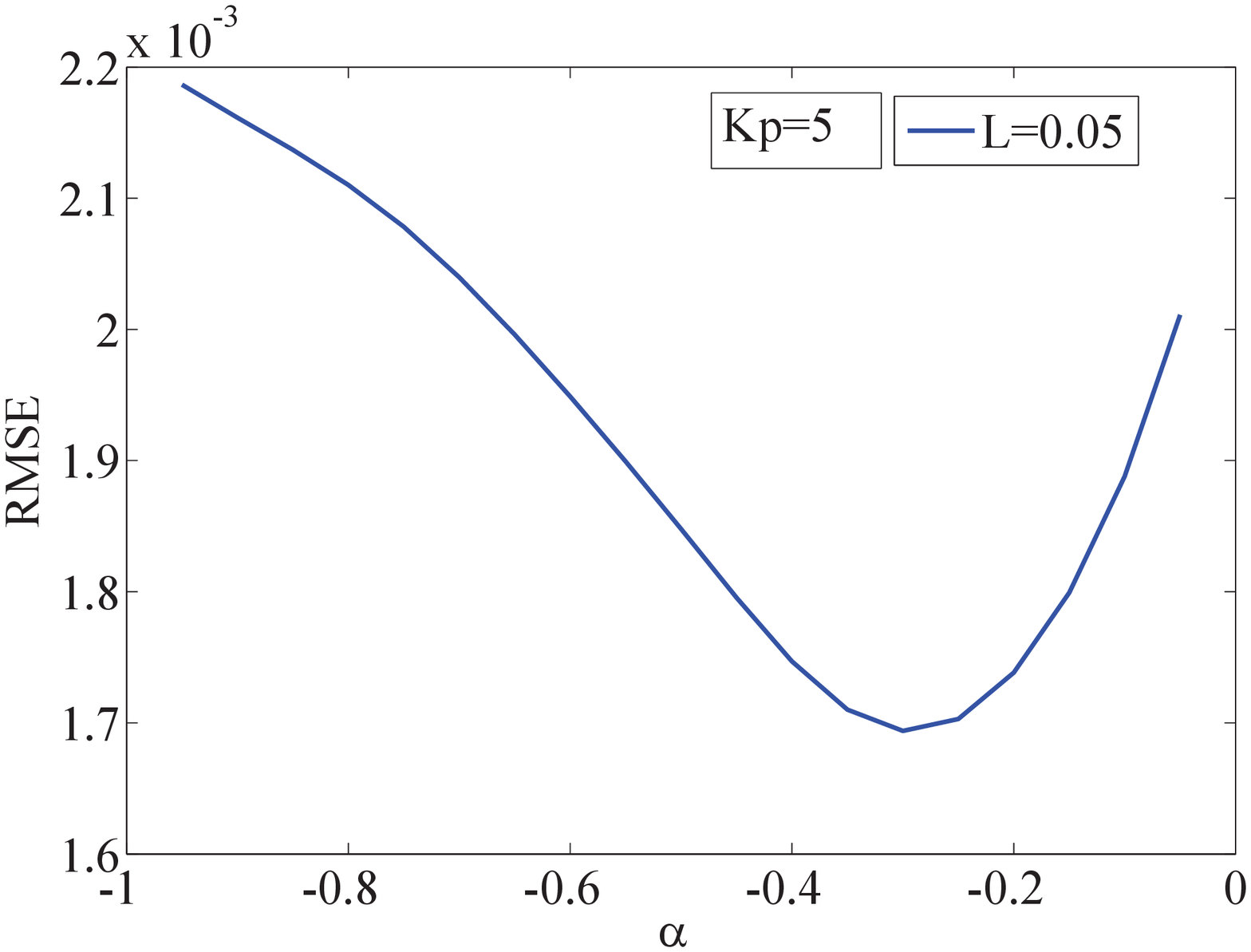}
  \caption{The RMSE scan result to different $\alpha$ with fixed $L$, and $K_p=5$}
  \label{fig10}
\end{figure}

\begin{figure}[!htb]
  \centering
  \includegraphics[width=0.9\hsize]{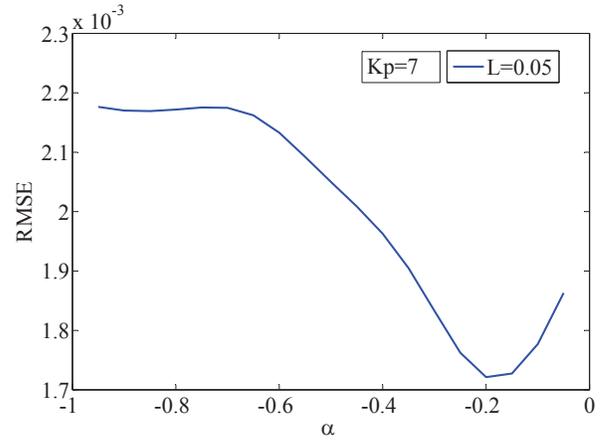}
  \caption{The RMSE scan result to different $\alpha$ with fixed $L$, and $K_p=7$}
  \label{fig11}
\end{figure}

~~~(3)~ When the gain of the human operator gets the value $K_p=5$ or $K_p=7$, the RMSE scan results for each $L$ are shown in Fig.(\ref{fig10}) and Fig.(\ref{fig11}) respectively. From the figures it can be seen that the corresponding $\alpha$ to the minimum RMSE is fractional. Meanwhile as the $K_p$ gets the big value in this two cases, Fig.(\ref{fig10}) and Fig.(\ref{fig11}) only show the RMSE to $L=0.05$, and the RMSE to other $L$ (which is greater than 0.05 seconds) is too large to be shown in the figures.
\section{Experiment Research}
\begin{figure}[!htb]
  \centering
  \includegraphics[width=0.85\hsize]{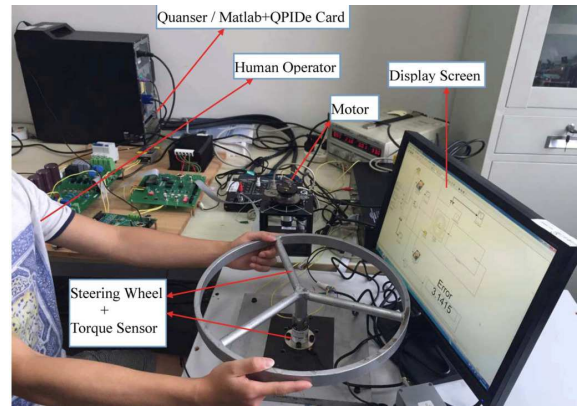}
  \caption{The human-in-the-loop control experiment platform}
  \label{fig12}
\end{figure}

\begin{figure}[!htb]
  \centering
  \includegraphics[width=0.85\hsize]{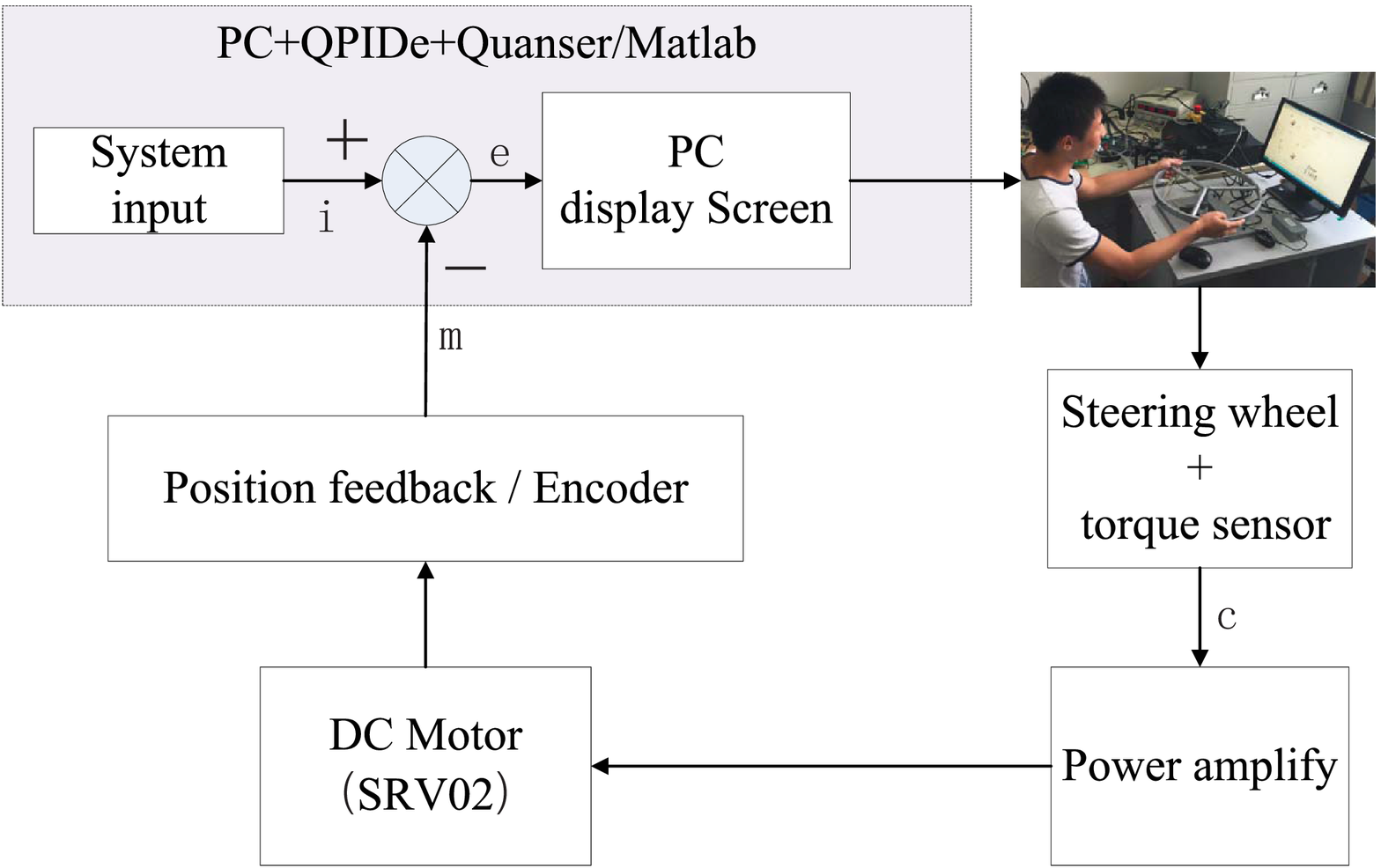}
  \caption{The block diagram of the human-in-the-loop control experiment }
  \label{fig13}
\end{figure}

In this section, the human-in-the-loop control experiment will be done based on the Quanser SRV02 Rotary Servo Base unit. The experiment platform is shown in Fig.\ref{fig12}, which is composed of a human operator, a steering wheel, a torque sensor, a motor, a computer installed with Quanser/Matlab real time software and QPIDe data acquisition card. The steering wheel is fixed with the torque sensor which is mounted on the desk. The voltage output of the torque sensor is power amplified and transferred to the motor. The motor works on voltage to position control mode, and the encoder on the motor offers a high resolution of 4096 counts per revolution in quadrature mode(1024 line per revolution). The QPIDe card samples the voltage output of the torque sensor together with the encoder output. In the experiment, the system input, output and error information are all shown on the display screen of the computer, and the human operator observes the system error and applies a force around the steering wheel, and so controls the motor's position to follow the system input. The block diagram of the human-in-the-loop control system is shown in Fig.\ref{fig13}.

\begin{figure}[!htb]
  \centering
  \includegraphics[width=0.9\hsize]{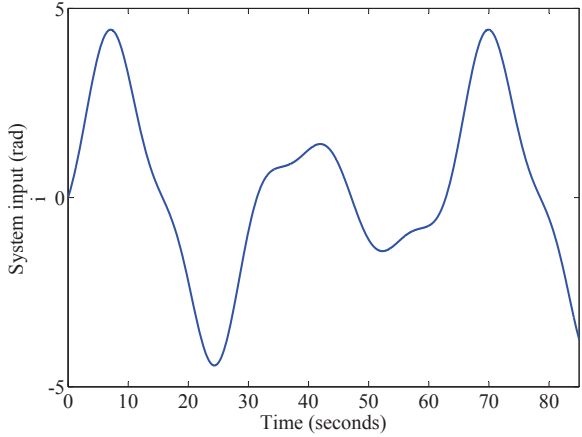}
  \caption{The system input of the human-in-the-loop control experiment}
  \label{fig14}
\end{figure}

\begin{figure}[!htb]
  \centering
  \includegraphics[width=0.9\hsize]{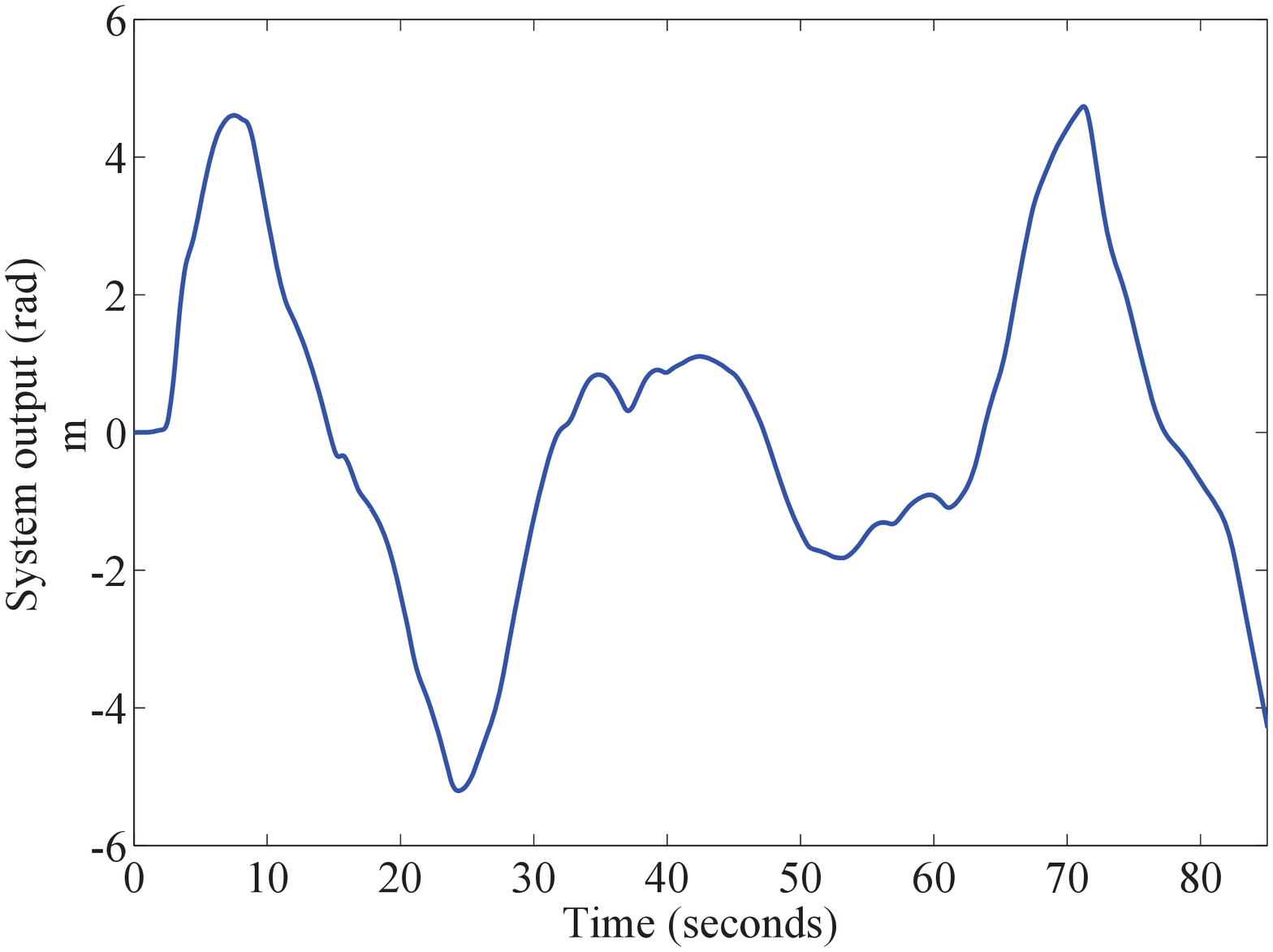}
  \caption{The system output of the human-in-the-loop control experiment}
  \label{fig15}
\end{figure}

\begin{figure}[!htb]
  \centering
  \includegraphics[width=0.9\hsize]{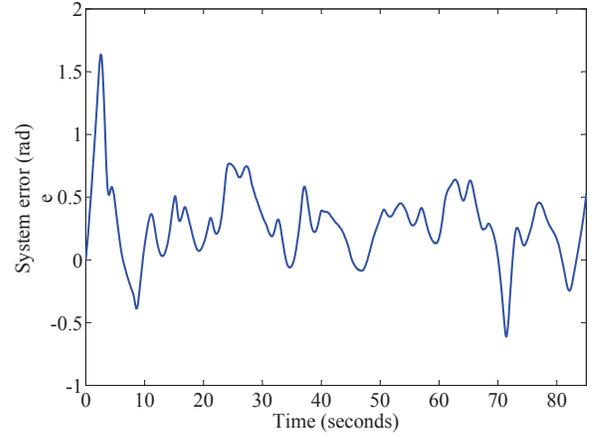}
  \caption{The system error of the human-in-the-loop control experiment}
  \label{fig16}
\end{figure}

\begin{figure}[!htb]
  \centering
  \includegraphics[width=1\hsize]{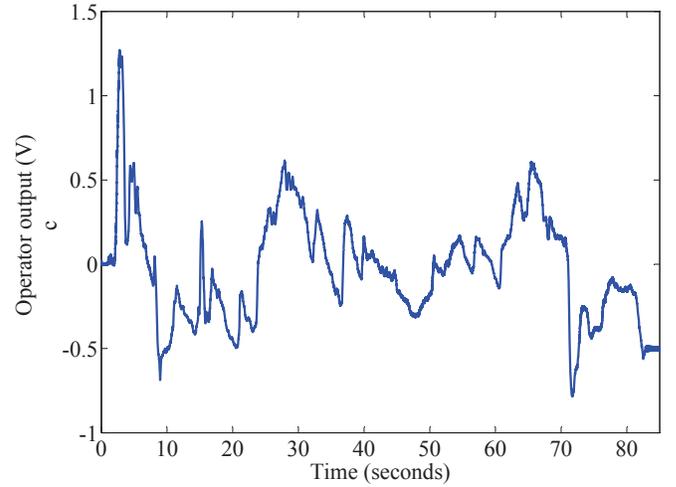}
  \caption{The human operator output(1V=4N.m)}
  \label{fig17}
\end{figure}

In the experiment, the motor works on position control mode, in this case it is a second order system and its transfer function  is described as follow:
\begin{equation}
\label{eq13}
Y_c({\rm{s}}) = \frac{K}{{s(\tau s + 1)}}{\rm{ = }}\frac{{{\rm{ 60}}{\rm{.2362}}}}{{s(s + 39.37)}},
\end{equation}
where $K=1.53\rm{rad/s/V}$, $\tau=0.0254s$. In this experiment, the time delay of the human operator's delay is tested about $L=0.3\rm{s}$, and the system input i(t), system output m(t), system error e(t) and operator output c(t) are real time recorded as shown in Fig.\ref{fig14} to Fig.\ref{fig17}.


\subsection{ The minimum RMSE and $best\_fit$ parameters for each models}

 Using the experiment data and the searching criteria Eq.(\ref{eq10})$-$  Eq.(\ref{eq12}), the minimum RMSE and the corresponding $best\_fit$ parameters value for each model are obtained as shown in Table \ref{table_2}. From Table \ref{table_2}, it is obvious that the proposed fractional order model described by Eq.(\ref{eq8}) has the smallest RMSE, and the corresponding order of the model is $\alpha=-0.3873$. This means that compared with the traditional model, the proposed fractional order model described by Eq.(\ref{eq8}) is the $best\_fit$ model for describing the human operator behavior, in other word, the human operator is a fractional order system. This result is consistent with the result obtained in section IV.
\begin{table}[h]
\caption{$best\_fit$ parameters value and RMSE for each model ($L=0.3\rm{s}$)}
\label{table_2}
\begin{center}
\begin{tabular}{|c|l|c|}

\hline
Model                                                                                              & Parameters   & Values  \\ \cline{2-3}

\hline
\multirow{4}*{$Y_{P3}(s){\rm{ = }}\frac{{K_p e^{ - Ls} }}{{s^\alpha  }}$}                           & RMSE        & $3.751\times 10^{-3}$ \\ \cline{2-3}
                                                                                                    & $\alpha^*$  &-0.3873 \\ \cline{2-3}
                                                                                                    & $K_p^*$     & 0.7643 \\ \cline{2-3}

\hline
\multirow{3}*{$Y_{P2}(s){\rm{ = }}K_p(s+3)e^{-Ls}$}                                                 & RMSE         & $4.172\times 10^{-3}$  \\ \cline{2-3}
                                                                                                    & $K_p^*$      & 0.6099   \\ \cline{2-3}
\hline
\multirow{6}*{$Y_{P1} {\rm{(}}s{\rm{)=}}\frac{{K_p (T_L s + 1)e^{ - Ls} }}{{(T_I s + 1)(T_N s + 1)}}$}  & RMSE     & $4.036\times 10^{-3}$   \\ \cline{2-3}
                                                                                                     & $K_p^*$     & 1.078   \\ \cline{2-3}
                                                                                                     & $T_L^*$     &0.1481    \\ \cline{2-3}
                                                                                                     & $T_I^*$     &0.0001      \\ \cline{2-3}
                                                                                                     & $T_N^*$     & 0.7804     \\ \cline{2-3}
\hline
\end{tabular}
\end{center}
\end{table}

\subsection{The models parameters to different $L$ }

\begin{figure}[!htb]
  \centering
  \includegraphics[width=1\hsize]{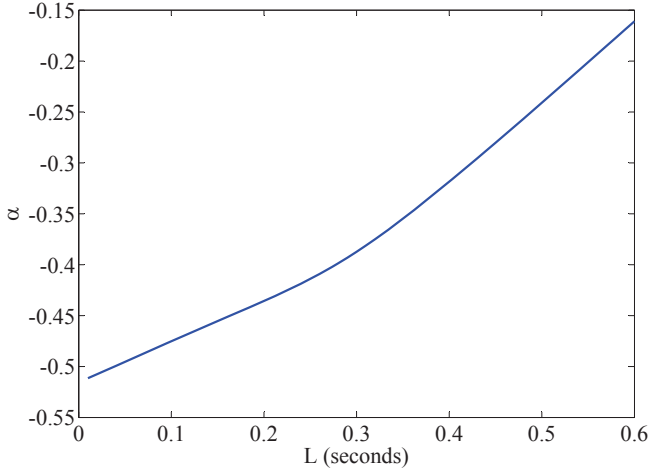}
  \caption{The fractional order $\alpha$ distribution of human operator to different $L$ using the proposed model described by Eq.(\ref{eq8})}
  \label{fig18}
\end{figure}

In general, the time delay of human operator varies in small range, so in this section the proposed fractional order model described by Eq.(\ref{eq8}) and the conventional model described by Eq.(\ref{eq5}) will be considered, and the models parameters distribution to different human time delay $L$ will be scanned. As the time delay of human operator has finite range, so in this scanning process the time delay $L$ varies from $0.01$ to $0.6$ with $0.01$ step length. The scan results are shown in Fig.(\ref{fig18}) to Fig.(\ref{fig21}).

\begin{figure}[!htb]
  \centering
  \includegraphics[width=1\hsize]{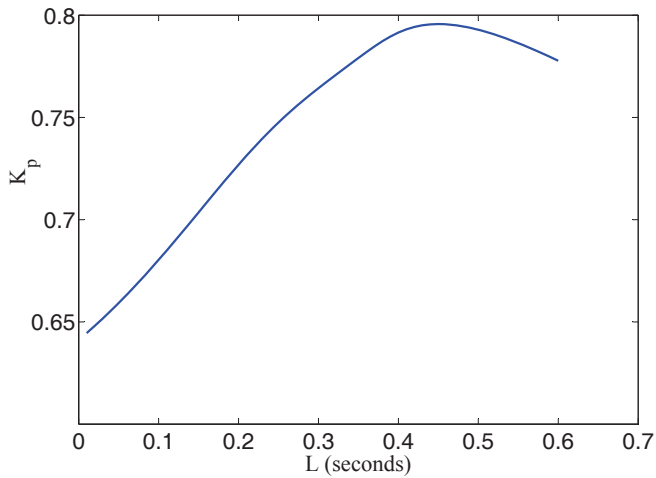}
  \caption{The gain $K_p$ distribution of human operator to different L using the proposed model described by Eq.(\ref{eq8})}
  \label{fig19}
\end{figure}
\begin{figure}[!htb]
  \centering
  \includegraphics[width=1\hsize]{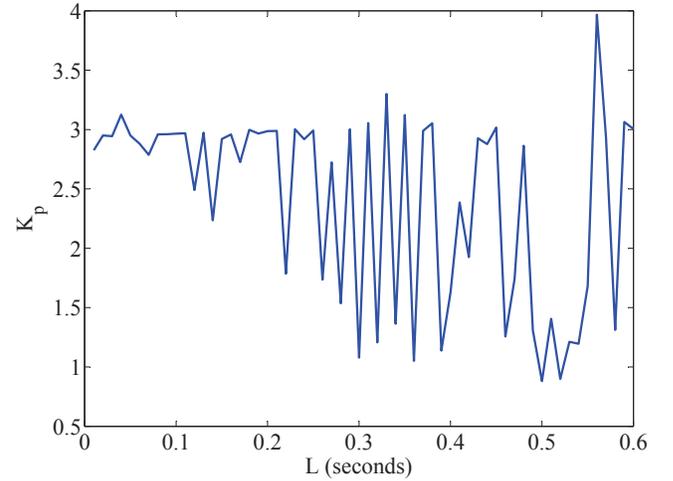}
  \caption{The gain $K_p$ distribution of human operator to different L using the conventional model described by Eq.(\ref{eq5})}
  \label{fig20}
\end{figure}
\begin{figure}[!htb]
  \centering
  \includegraphics[width=1\hsize]{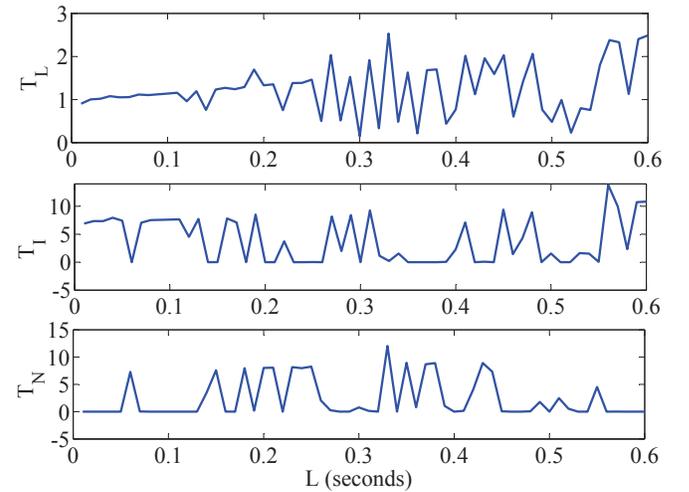}
  \caption{The $T_L$,$T_I$, $T_N$ distributions of human operator to different L using the conventional model described by Eq.(\ref{eq5})}
  \label{fig21}
\end{figure}

Fig.(\ref{fig18}) and Fig.(\ref{fig19}) show that the distributions of $\alpha$ and $K_p$ of the proposed fraction order model are smooth, meanwhile as the time delay $L$ decreases, the fractional order $\alpha$ tend to negative increase. Fig.(\ref{fig20}) and Fig.(\ref{fig21}) show that the parameters $K_p$, $T_L$, $T_I$ and $T_N$ of the conventional model described by Eq.(\ref{eq5}) fluctuate in large scale. From this point of view, the proposed fractional order model described by Eq.(\ref{eq8}) is suitable to describe the human operator behavior.

\section{Conclusions}

In this paper, based on the characteristics of human brain and behaviour, the fractional order mathematical model for human operator is proposed. Based on the actual data, the models verifications have been done, and the $best\_fit$ parameters for the proposed model and the traditional models have been obtained. The verification results show that the proposed fractional order model described by Eq.(\ref{eq8}) is the $best\_fit$ model for describing the human operator behavior, in other words, the human operator is a fractional order in such a system. The experiment results also provide the correctness of the above conclusion.

The proposed fractional order model described by Eq.(\ref{eq8}) for human operator behavior not only has small RMSE, but also has a simple structure with only few parameters, and each parameter has definite physical meaning.

In the future work, we will research the model for human operator considering other types of controlled element.

\section*{ACKNOWLEDGMENT}

I would like to thank the graduate student Cui Lei who helped in digitizing the old data.
\ifCLASSOPTIONcaptionsoff
  \newpage
\fi






\begin{thebibliography}{99}

\bibitem{c1} Tustin A, The nature of the operator's response in manual control, and its implications for controller design, \emph{Journal of the Institution of Electrical Engineers-Part IIA: Automatic Regulators and Servo Mechanisms}, vol. 94, no. 2, pp. 190-206, 1947.

\bibitem{c2} Craik K J W, Theory of the human operator in control systems, \emph{British Journal of Psychology, General Section}, vol. 38, no. 3, pp. 142-148, 1948.

\bibitem{c3} McRuer D T, Krendel E S, The human operator as a servo system element, \emph{Journal of the Franklin Institute}, vol. 267, no. 6, pp. 511-536, 1959.

\bibitem{c4}  Roig R W, A comparison between human operator and optimum linear controller RMS-error performance, \emph{IRE Transactions on Human Factors in Electronics}, vol. 1, pp. 18-21, 1962.

\bibitem{c5} Senders J W, The human operator as a monitor and controller of multidegree of freedom systems, \emph{IEEE Trans. Human Factors in Electronics}, vol. 1, pp. 2-5, 1964.

\bibitem{c6} McRuer D T, Human operator dynamics in compensatory systems, \emph{SYSTEMS TECHNOLOGY INC HAWTHORNE CA}, 1965.

\bibitem{c7} McRuer D T, Jex H R, A review of quasi-linear pilot models, \emph{IEEE Trans. Human Factors in Electronics}, vol. 3, pp. 231-249, 1967.

\bibitem{c8} McRuer D T, Hofmann L G, Jex H R, et al, New approaches to human-pilot/vehicle dynamic analysis, \emph{SYSTEMS TECHNOLOGY INC HAWTHORNE CA}, 1968.

\bibitem{c12} Baron S, Kleinman D L, The human as an optimal controller and information processor, \emph{IEEE Trans. Man-Machine Systems}, vol. 10, no. 1, pp. 9-17, 1969.

\bibitem{c14} Phatak A V, Bekey G A, Model of the adaptive behavior of the human operator in response to a sudden change in the control situation, \emph{IEEE Trans. Man-Machine Systems}, vol. 10, no. 3, pp. 72-80, 1969.

\bibitem{c15} Wierenga R D, An evaluation of a pilot model based on Kalman filtering and optimal control, \emph{IEEE Trans. Man-Machine Systems}, vol. 10, no. 4, pp. 108-117, 1969.

\bibitem{c16} Levison W H, Baron S, Kleinman D L, A model for human controller remnant, \emph{IEEE Trans. Man-Machine Systems}, vol. 10, no. 4, pp. 101-108, 1969.

\bibitem{c17} Kleinman D L, Baron S, Levison W H, An optimal control model of human response part I: Theory and validation, \emph{Automatica}, vol. 6, no. 3, pp. 357-369, 1970.

\bibitem{c18} Baron S, Kleinman D L, Levison W H, An optimal control model of human response part II: prediction of human performance in a complex task, \emph{Automatica}, vol. 6, no, 3, pp. 371-383, 1970.

\bibitem{c22} McRuer D T, Krendel E S, Mathematical models of human pilot behavior, \emph{ADVISORY GROUP FOR AEROSPACE RESEARCH AND DEVELOPMENT NEUILLY-SUR-SEINE (FRANCE)}, 1974.

\bibitem{c24}  Tomizuka M, Whitney D E, The human operator in manual preview tracking (an experiment and its modeling via optimal control), \emph{Journal of Dynamic Systems, Measurement, and Control}, vol. 98, no. 4, pp. 407-413, 1976.

\bibitem{c25}  Phatak A, Weinert H, Segall I, et al, Identification of a modified optimal control model for the human operator, {Automatica}, vol. 12, no.1, pp. 31-41, 1976.

\bibitem{c27} Gabay E, Merhav S J, Identification of a parametric model of the human operator in closed-loop control tasks, \emph{IEEE Trans. Systems, Man and Cybernetics}, vol. 7, no. 4, pp. 284-292, 1977.

\bibitem{c30} McRuer D, Human dynamics in man-machine systems, {Automatica}, vol. 16, no. 3, pp. 237-253, 1980.

\bibitem{c36} Govindaraj T, Ward S L, Poturalski R J, et al, An experiment and a model for the human operator in a time-constrained competing-task environment, \emph{IEEE Trans. Systems, Man and Cybernetics}, vol.4, 496-503, 1985.

\bibitem{c39} Sworder D, Haaland K S, A hypothesis evaluation model for human operators, \emph{IEEE Trans. Systems, Man and Cybernetics}, vol. 19, no. 5, pp. 1091-1100, 1989.

\bibitem{c50}  Boer E R, Kenyon R V, Estimation of time-varying delay time in nonstationary linear systems: an approach to monitor human operator adaptation in manual tracking tasks, \emph{IEEE Trans. Systems, Man and Cybernetics, Part A: Systems and Humans}, vol. 28, no. 1, pp. 89-99, 1998.

\bibitem{c51} Phillips C A, Repperger D W, An informatic model of human operator control, \emph{[Engineering in Medicine and Biology, 1999. 21st Annual Conference and the 1999 Annual Fall Meetring of the Biomedical Engineering Society] BMES/EMBS Conference, 1999. Proceedings of the First Joint. IEEE}, 2: 1009 vol. 2, 1999.

\bibitem{c53} Doman D B, Anderson M R, A fixed-order optimal control model of human operator response, \emph{Automatica}, vol. 36, no. 3, pp. 409-418, 2000.

\bibitem{c57} Macadam C C, Understanding and modeling the human driver, \emph{Vehicle System Dynamics}, vol. 40, no.(1-3), pp. 101-134, 2003.

\addtolength{\textheight}{-15cm}   


\bibitem{c64} Kovacevic D, Pribacic N, Jovic M, et al, Modeling Human Operator Controlling Process in Different Environments, \emph{Artificial Neural Networks¨CICANN} 2009. Springer Berlin Heidelberg, pp. 475-484, 2009.

\bibitem{c68} Celik O, Ertugrul S, Predictive human operator model to be utilized as a controller using linear, neuro-fuzzy and fuzzy-ARX modeling techniques. \emph{Engineering Applications of Artificial Intelligence}, vol. 23, no. 4, pp. 595-603, 2010.

\bibitem{c69} Tervo K, Discrete data-based state feedback model of human operator, \emph{Mechatronics and Embedded Systems and Applications (MESA), 2010 IEEE/ASME International Conference on. IEEE}, pp. 202-207, 2010.


\bibitem{c71}  Zaal P M T, Sweet B T, Estimation of time-varying pilot model parameters, \emph{AIAA Paper}, 2011, 6474: 2011.

\bibitem{c88}  Zhang B, Li H Y, Tang G J, Human control model in teleoperation rendezvous, \emph{Science China Information Sciences}, pp. 1-11, 2014.

\bibitem{c89}  Lone M, Cooke A, Review of pilot models used in aircraft flight dynamics, \emph{Aerospace Science and Technology}, vol. 34, pp. 55-74, 2014.

\bibitem{c91} Li W, Sadigh D, Sastry S S, et al, Synthesis for Human-in-the-Loop Control Systems, \emph{Tools and Algorithms for the Construction and Analysis of Systems}, Springer Berlin Heidelberg, pp. 470-484, 2014.

\bibitem{c95} Rao M S P, The human operator in man-machine systems, \emph{Defence Science Journal}, vol. 6, no. 3, pp. 182-190, 2014.

\bibitem{c96}  Liu Y K, Zhang Y M, Control of human arm movement in machine-human cooperative welding process, \emph{Control Engineering Practice}, vol. 32, pp. 161-171, 2014.


\bibitem{Chen2009}
Chen Y Q, Petras I, Xue D Y, Fractional order control-A tutorial, \emph{American Control Conference}, pp.1397-1411, 2009.

\bibitem{Chen2010}
Li Y, Chen Y Q, Podlubny I, Stability of fractional order nonlinear dynamic systems: Lyapunov direct method and generalized Mittag-Leffler stability, \emph{Computers Mathematics with Applications}, vol. 59, no. 5, pp. 1810-1821, 2010.

\bibitem{Oldham1974}
Oldham K B, Spanier J, The fractional calculus, \emph{Academic Press}, New York, 1974.

\bibitem{Podlubny1999}
Podlubny I, Fractional-order systems and $PI^\lambda D^u$- controllers, \emph{IEEE Trans. Automatic Control}, vol. 44, no. 1, pp. 208-214,1999.

\bibitem{Li2010}
Li H S, Luo Y, Chen Y Q, A fractional order proportional and derivative (FOPD) motion controller: tuning rule and experiments, \emph{IEEE Trans. Control Systems Technology}, vol. 18, no. 2, pp. 516-520, 2010.

\bibitem{Podlubny1999-2}
Podlubny I, Fractional differential equations, vol. 198, \emph{Academic Press}, San Diego, Calif, USA, 1999.

\end{thebibliography}
\end{document}